\documentclass[9pt,conference,a4paper]{IEEEtran}
\usepackage{cite}

\ifCLASSINFOpdf
   \usepackage[pdftex]{graphicx}
\else
\fi

\begin{document}

\title{The varying $w$ spread spectrum effect \\for radio interferometric imaging}

\author{%
\IEEEauthorblockN{
Laura Wolz\IEEEauthorrefmark{1}\IEEEauthorrefmark{2}, 
Filipe B. Abdalla\IEEEauthorrefmark{1}, 
Rafael E. Carrillo\IEEEauthorrefmark{3}, 
Yves Wiaux\IEEEauthorrefmark{3}\IEEEauthorrefmark{4}\IEEEauthorrefmark{5},
Jason D. McEwen\IEEEauthorrefmark{1}}
\IEEEauthorblockA{\IEEEauthorrefmark{1} 
Department of Physics and Astronomy, 
University College London, 
London WC1E 6BT, 
UK}
\IEEEauthorblockA{\IEEEauthorrefmark{2}
Sub-Dept. of Astrophysics, Dept. of Physics, University of Oxford,
The Denys Wilkinson Building,
Keble Road,
Oxford
OX1 3RH,
UK}
\IEEEauthorblockA{\IEEEauthorrefmark{3}
Institute of Electrical Engineering, 
Ecole Polytechnique Federale de Lausanne (EPFL), 
CH-1015 Lausanne, 
Switzerland}
\IEEEauthorblockA{\IEEEauthorrefmark{4}
Department of Radiology and Medical Informatics, 
University of Geneva (UniGE), CH-1211 Geneva, 
Switzerland}
\IEEEauthorblockA{\IEEEauthorrefmark{5}
Department of Radiology, 
Lausanne University Hospital (CHUV), CH-1011 Lausanne, Switzerland}
}

\maketitle

\begin{abstract}
We study the impact of the spread spectrum effect in radio interferometry on the quality of image reconstruction.  This spread spectrum effect will be induced by the wide field-of-view of forthcoming radio interferometric telescopes. The resulting chirp modulation improves the quality of reconstructed interferometric images by increasing the incoherence of the measurement and sparsity dictionaries.  We extend previous studies of this effect to consider the more realistic setting where the chirp modulation varies for each visibility measurement made by the telescope. In these first preliminary results, we show that for this setting the quality of reconstruction improves significantly over the case without chirp modulation and achieves almost the reconstruction quality of the case of maximal, constant chirp modulation.
\end{abstract}


\section{Introduction}

The next generation of radio interferometers will see a large field of view. Consequently, the planar interferometric imaging setting considered typically needs to be adapted to a wide field-of-view by incorporating the so-called $w$-term component, introducing a chirp modulation.  

\section{Compressed Sensing in Radio Interferometry}

Previous studies have shown the power of the compressed sensing formalism in radio interferometric imaging \cite{2009MNRAS.395.1733W, 2012MNRAS.426.1223C, 2009MNRAS.400.1029W, 2011MNRAS.413.1318M}.
Radio interferometers acquire incomplete Fourier measurements, so-called visibilities, of the image on the sky under observation. Recovering an image from the visibilities measured by the telescope is hence an ill-posed inverse problem, which is solved through convex optimisation methods (e.g.\ \cite{2009MNRAS.395.1733W, 2009MNRAS.400.1029W}). 

A crucial factor controlling the fidelity of reconstruction in this approach is the incoherence of the measurement and sparsity dictionaries.  In the wide field-of-view setting, the chirp modulation that is induced acts to increase incoherence.  For radio interferometry, the measurement basis can essentially be identified with the Fourier basis.  In this case the coherence is given by the maximum modulus of the Fourier coefficient of the sparsity atoms.  The chirp modulation corresponds to a norm-preserving convolution in Fourier space, spreading the spectrum of the sparsity atoms, thus reducing the maximum modulus of their Fourier coefficients and increasing incoherence.  The increased incoherence due to this spread spectrum effect acts to improve the fidelity of image reconstruction \cite{2009MNRAS.400.1029W}.

\section{First Results and Outlook}

In this preliminary work we first confirm previous results \cite{2009MNRAS.400.1029W} that demonstrate the effectiveness of the spread spectrum phenomenon, however here we consider more realistic interferometric images and alternative sparsity dictionaries. We then extend the constant chirp analysis of previous studies \cite{2009MNRAS.400.1029W, 2011MNRAS.413.1318M} to the more realistic setting where every measurement in the $(u,v)$-space of visibilities undergoes a different $w$-term modulation.  This is a computationally demanding setting which we address by incorporating the $w$-projection algorithm \cite{2008ISTSP...2..647C} into our framework.  We consider uniform visibility sampling in $(u,v)$ and $w$, with $w$ samples ranging from zero to $2/L$ times the maximum values of $u$ and $v$, where $L$ corresponds to the size of the field-of-view (identical to the maximum $w$ considered previously \cite{2009MNRAS.400.1029W}).  We denote the $w$ range by the discrete component $0<w_d<1$ respectively.  Hence, $w_d=0$ corresponds to no chirp modulation, $w_d=1$ corresponds to the maximal chirp modulation studied previously \cite{2009MNRAS.400.1029W}, and the range $0<w_d<1$ corresponds to uniformly random sampling over the entire range.

\begin{figure}
\centering
\includegraphics[width=0.44\textwidth, clip=true, trim= 30 180 50 200]{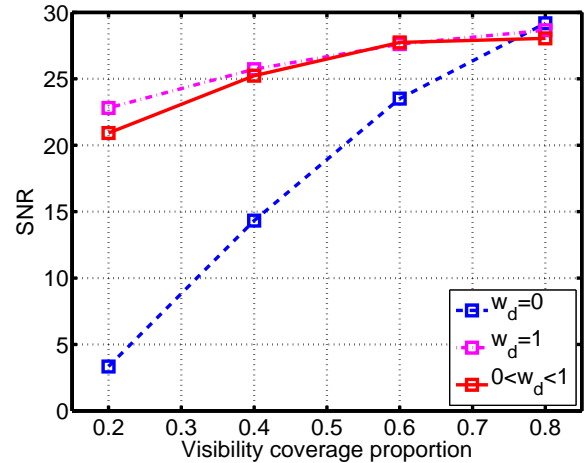}
\caption{Signal-to-noise ratio of the recovered image of M31 with Daubechies 8 wavelets for no chirp (blue, dashed), a constant maximal chirp (magenta, dash-dotted) and a varying $w$-modulation (red, solid) as a function of visibility coverage.}
\label{fig1}
\end{figure}

Preliminary results on a small test image of an HII region of M31 (see e.g.~\cite{2012MNRAS.426.1223C}) of dimension $120\times120$ pixels show that reconstruction fidelity is significantly improved compared to the analysis without chirp modulation when using Daubechies 8 wavelets (see Fig.~\ref{fig1}), extending previous findings to more realistic images and alternative sparsity dictionaries.  Furthermore, Fig.~\ref{fig1} shows that reconstruction fidelity for the varying $w$ case is almost as good as the constant, maximal chirp modulation. The study of the varying $w$ spread spectrum effect in the context of the SARA algorithm \cite{2012MNRAS.426.1223C} is ongoing.

\bibliographystyle{IEEEtran}
\bibliography{bib}

\end{document}